\documentclass[12pt]{article}
\usepackage{amssymb}
\usepackage{amsmath}
\usepackage[dvips]{graphicx}
\usepackage{epsfig}

\usepackage{ytableau}
\newcommand{\ket}[1]{\left|#1\right>}
\newcommand{\be}{\begin{equation}}
\newcommand{\ee}{\end{equation}}

\begin{document}

\baselineskip=18pt
\numberwithin{equation}{section}
\allowdisplaybreaks  

\pagestyle{myheadings}

\thispagestyle{empty}

\vspace*{0.5cm}
\begin{center}
 {\LARGE   \bf A non-renormalization theorem for chiral primary 3-point functions\\}
 \vspace*{1.7cm}
Marco Baggio$^{\,a}$, Jan de Boer$^{\,a}$ and Kyriakos Papadodimas$^{\,b}$ \\

 \vspace*{1.0cm}
   $^{a}$ Institute for Theoretical Physics, University of Amsterdam\\
Science Park 904, Postbus 94485, 1090 GL Amsterdam, The Netherlands\\
\vspace*{0.4cm}
$^{b}$ Theory Group, Physics Department, CERN, CH-1211 Geneva 23, Switzerland\\
 \vspace*{0.8cm}
 {\tt m.baggio@uva.nl, j.deboer@uva.nl, Kyriakos.Papadodimas@cern.ch}
\end{center}
\vspace*{1.0cm}

\noindent In this note we prove a non-renormalization theorem for the
3-point functions of 1/2 BPS primaries in the four-dimensional ${\cal
  N}=4$ SYM and chiral primaries in two dimensional ${\cal N}=(4,4)$
SCFTs.  Our proof is rather elementary: it is based on Ward identities
and the structure of the short multiplets of the superconformal
algebra and it does not rely on superspace techniques. We also discuss some possible generalizations to less supersymmetric multiplets.

\newpage
\setcounter{page}{2}

\tableofcontents

\section{Introduction}

\vskip20pt

One of the earliest checks of the AdS/CFT correspondence
\cite{Maldacena:1997re, hep-th/9802109, hep-th/9802150} was the
matching of 3-point functions of chiral primaries. This was first done
\cite{hep-th/9806074,D'Hoker:1998tz, D'Hoker:1999ea} for the duality between the ${\cal N}=4$ SYM and
IIB string theory in AdS$_5\times$S$^5$ and later
\cite{hep-th/0703001, hep-th/0703022, Pakman:2007hn, Taylor:2007hs}
for the duality between the two dimensional ${\cal N}=(4,4)$ D1/D5 CFT
and IIB string theory on AdS$_3\times$S$^3\times {\cal M}_4$. The
matching of 3-point functions is non-trivial because they are not
fully determined by symmetry considerations.

Notice that a priori the matching did not have to work --- i.e. even
if it did not work, it would not indicate a problem with the AdS/CFT
correspondence. The bulk and boundary computations of 3-point
functions are performed at different points of the moduli
space (i.e. different values of the coupling constants). In general
there is no reason to expect that such computations should give the
same answer.  The fact that the computations did indeed agree, strongly
suggests that these 3-point functions are actually independent of the
coupling constant. In other words, that there should exist a
``non-renormalization theorem" for 3-point functions of chiral
primaries in superconformal field theories with sufficient amount of
supersymmetry.

For the case of AdS$_5$/CFT$_4$ and the ${\cal N}=4$ SYM a proof of
such a non-renormalization theorem was given in a series of works
\cite{Intriligator:1998ig, Intriligator:1999ff, Eden:1999gh, Petkou:1999fv,
  Howe:1999hz, Heslop:2001gp}. The proof relies on the formalism of
analytic superspace. It would be interesting if a more basic
proof could be found, which would not depend on the somewhat elaborate construction
of analytic superspace\footnote{In \cite{Basu:2004nt} such a proof was proposed. However we believe that the arguments in that paper are  actually not sufficient in order to prove the non-renormalization theorem. More explanations about this can be found at the end of section 5.1, in particular see footnote 11.}.

In the case of AdS$_3$/CFT$_2$ with ${\cal N}=(4,4)$ supersymmetry a 
(partial) non-renormalization theorem was proven in
\cite{arXiv:0809.0507} using elementary techniques. This theorem is
partial because it does not include the most general case of 3-point
function of chiral primaries, but only the case of ``extremal"
correlation functions.

The goal of this work is two-fold. First we complete the
non-renormalization theorem of \cite{arXiv:0809.0507} to include the
most general 3-point function of chiral primaries in two-dimensional
${\cal N}=(4,4)$ theories.\footnote{In this paper we call ``chiral primaries'' all the operators belonging to the same $SU(2)$ R-symmetry multiplet, and not just the highest-weight state.} Second, we give a short --- and in our view
simpler --- proof of the non-renormalization theorem for 1/2 BPS chiral
primary 3-point functions for ${\cal N}=4$ SCFTs in four
dimensions. Our presentation provides a unified treatment of both
cases, based on superconformal Ward identities and the structure of
the representations of the superconformal algebra.

We also prove a few more results: 

i) $\,\,$ 3-point functions of half-chiral primary states in 2d ${\cal
  N}=(4,4)$ SCFTs are not renormalized

ii) $\,\,$ 3-point functions of chiral primaries in 2d ${\cal N}=(0,4)$ SCFTs are not
renormalized.

iii) ``Extremal'' $n$-point functions of 1/2 BPS operators in 4d ${\cal
  N}=4$ SCFTs are not renormalized

iv) $\,$ 3-point functions involving one 1/4 BPS and two 1/2 BPS operators
in 4d ${\cal N}=4$ SCFTs are not renormalized.

Notice that our results are non-perturbative in the coupling constant
of the theory and hold for any gauge group --- in particular they do not depend on a large $N$ limit.

The plan of the paper is as follows: in section 2 we present some
necessary background material mostly on marginal deformations of CFTs,
Ward identities, the structure of short multiplets and their 3-point
functions. In section 3 we outline the main proof of the
non-renormalization theorem in general context. In section 4 we
provide a detailed proof of the theorem for 2d ${\cal N}=(4,4)$
SCFTs. In section 5 we present a detailed proof of the theorem for 4d
${\cal N}=4$ theories. In the remaining sections and appendices we
provide various additional details.

\section{Background material}

In this section we review the basic ingredients that go into the proof
of the non-renormalization theorem. The reader who is familiar with
basic properties of superconformal field theories can skip directly
to the next section.

\subsection{Conformal perturbation theory}

Our goal is to understand the coupling constant dependence of certain
correlation functions. Changing a coupling constant $g$ in a CFT
corresponds to deforming the CFT by an exactly marginal operator
${\cal O}$. Correlators in the deformed theory can be computed from ${\it integrated}$ correlators in the undeformed CFT. We have schematically
\be
\label{schd}
\frac{\partial}{\partial g}\langle{\cal O}_1(x_1)\ldots
{\cal O}_n(x_n)\rangle \sim \int d^d x\,\langle{\cal O}(x) {\cal O}_1(x_1)\ldots
{\cal O}_n(x_n)\rangle
\ee This is only schematic because the integral has to be regularized
due to UV divergences when $x$ approaches the other
insertions. Because of these divergences and the need for
regularization, marginal deformations at second order do not commute,
we refer the reader to \cite{hep-th/9304053,
  arXiv:0809.0507,arXiv:0910.4963} for more details\footnote{\label{fn:ct}An
  alternative approach is to attribute this phenomenon to the presence
  of ``contact terms'', as explained in
  \cite{Seiberg:1988pf,Kutasov:1988xb}. Instead, the point of view we
  are adopting is that CFT correlators are only defined at distinct
  points and hence ``contact terms'' play absolutely no role. From
  this point of view operator mixing comes from the {\it definition}
  of the regularized integrated correlators, as was nicely discussed
  in \cite{hep-th/9304053}. The two approaches are equivalent, but we
  find it conceptually more clear to follow \cite{hep-th/9304053} and
  to avoid talking about contact terms. Hence, in the entirety of this
  paper we will never bring two local operators on the same spacetime
  point. }. Physically this can be understood as a certain kind of
operator mixing: under marginal deformations there is an ambiguity of
coupling-constant dependent redefinitions of operators with the same
quantum numbers.

The picture that we should keep in mind is that in general the moduli
space ${\cal M}$ (i.e. the space of marginal couplings of the CFT ---
sometimes called the ``conformal manifold'') is a higher
dimensional manifold and the local operators of the CFT are 
sections of vector bundles over ${\cal M}$.  So more precisely
instead of \eqref{schd}, what we have is that
\begin{equation}
\label{basicmarginal}
\nabla_g \langle{\cal O}_1(x_1)\ldots
{\cal O}_n(x_n)\rangle \equiv \int d^d x\,\langle{\cal O}(x) {\cal O}_1(x_1)\ldots
{\cal O}_n(x_n)\rangle
\end{equation}
In general $[\nabla_{g_1},\nabla_{g_2}]\neq 0$,
which expresses that there is non-trivial operator mixing over the
moduli space. The bundles on which operators take values have have non-trivial
connection which enters this covariant derivative.

In this paper we will prove that certain correlation
functions of chiral primaries do not depend on the couplings of the
CFT. More precisely, what we need to show is that the covariant
derivative of such correlators with respect to the couplings is
zero. This is the ``covariant'' way to phrase the non-renormalization
of correlation functions, which is unambiguous with respect to
coupling constant dependent operator redefinitions.

Actually we will prove a stronger statement. We will not only
show that --- in certain supersymmetric CFTs, and for specific choices
of the operators ${\cal O}_1\ldots{\cal O}_n$ --- the RHS of
\eqref{basicmarginal} vanishes, but we will show that the
      {\it integrand} on the RHS of \eqref{basicmarginal} is zero.
      This is a sufficient condition for the LHS to vanish. The
      integral is supposed to be carefully regularized, and the operators are
      never brought on top of each other, so there is no subtlety with possible
      ``contact terms'' (see also footnote \ref{fn:ct}). 

Let us then emphasize once more that if
\begin{equation}
\label{toshow}
 \langle {\cal O}(x) {\cal O}_1(x_1)\ldots{\cal O}_n(x_n) \rangle = 0
\end{equation} 
for distinct points, then it is guaranteed that the correlator
$\langle {\cal O}_1(x_1)\ldots{\cal O}_n(x_n) \rangle$ does not change
under marginal deformations by ${\cal O}$.

\subsection{Superconformal Ward identities}

For a general strongly coupled CFT there is no reason to expect
the vanishing of a correlator of the form \eqref{toshow}. The simplest
reason for a correlator to exactly vanish is because of a 
symmetry of the theory. For example, if the CFT has an (unbroken)
global $U(1)$ symmetry, then a correlator is automatically zero if the
charges of the inserted operators do not satisfy $\sum q_i = 0$. In a sense, our
proof will be based on similar conservation conditions, coming from the
supersymmetric (and superconformal) charges of the theory.

Symmetries in CFTs are expressed in terms of Ward identities. In the
case of a global internal symmetry with a conserved current $J_\mu$ we define the charge as $R = \int
d^{d-1}x\, J_0(x)$ and then we can show that for any set of local
operators we have
$$
\sum_{i=1}^n \langle {\cal O}_1(x_1)\ldots[R,{\cal O}_i](x_i)\ldots{\cal O}_n(x_n)\rangle = 0
$$ For global internal symmetries, this is the only type of Ward
identity that we have.

The situation is richer for conserved ``currents'' with additional
spacetime indices. For example, let us consider the stress energy
tensor which satisfies $\partial^\mu T_{\mu\nu}=0$. Consider an
arbitrary vector field $V_\mu(x)$ and construct the operator $
j^V_\mu(x) = V^\nu(x) T_{\mu\nu}(x)$. Using that
$T_{\mu\nu}$ is conserved and symmetric we have that $ \partial^\mu
j^V_\mu(x) = \frac12 (\partial^\mu V^\nu + \partial^\nu V^\mu)
T_{\mu\nu} $. Combining this with the tracelessness of $T_{\mu\nu}$ we
conclude that any vector field which satisfies $ \partial^\mu V^\nu +
\partial^\nu V^\mu = \omega(x) g^{\mu\nu} $ leads to a conserved
current $j^V_\mu$. Of course this is the condition for a conformal
Killing vector field. Provided that $V_\mu(x)$ does not grow too fast at
infinity, this can be used to define corresponding charges $R^V =
\int d^{d-1}x j^V_0(x)$ and corresponding Ward identities,
characterized by the choice of $V$. These conformal Ward identities are slightly
more complicated than the ones for global internal symmetries, but are of course
very well understood.

In this paper we will mostly use the superconformal Ward identities,
i.e. the identities that follow from the existence of a supercurrent
operator in the CFT. This is an operator of dimension $d-\frac12$
and two Lorentz indices, a vector index $\mu$ and a spinor index
$a$. Let us denote this operator as $G_{\mu a}$. We can construct (fermionic) conserved currents out of the
supercurrent by contracting it with a spinor valued field $\psi^a(x)$
as
$$
j^\psi_\mu(x) = \psi^a(x) G_{\mu a}(x)
$$ The condition for $j^\psi_\mu$ to be conserved is that $\psi^a(x)$ must
be a conformal Killing spinor. We also need to impose that it does not
grow too fast as $|x|\rightarrow \infty$ in order for the
corresponding charge $\int d^{d-1}x \,\,j^\psi_0(x)$ to be
well-defined. Then we find the following possibilities. The first
possibility is to take $\psi^a(x)$ to be a constant spinor independent of
$x$. Then the charges $\int j^\psi_0$ are the usual supercharges that
we denote by ${\bf Q}$. The second possibility is to take $\psi^a(x)$ to
be linear in $x$ and then the corresponding charges turn out to be the
``superconformal partners'' of ${\bf Q}$ that we denote by ${\bf S}$. \footnote{Our
  notation in this section is rather loose. By ${\bf S}$ we simply mean the
  ``superconformal partner'' of ${\bf Q}$ in the sense that they both
  come from the same supercurrent. In two dimensional notation we
  would have that ${\bf Q} \sim G_{-\frac12}$ while ${\bf S} \sim G_{+\frac12}$. In 4d SCFTs if by ${\bf Q}$ we denote one of
  the left-chiral supercharges $Q_a$ then the corresponding ${\bf S}$
  which comes from the same supercurrent is right-chiral ${\bf S} \sim
  \overline{S}_{\dot{a}}$. We hope the notation is not too
  confusing --- more details on the superconformal Ward identities for 4d SCFTs can be found in \cite{Basu:2004nt, arXiv:0910.4963}.
} For a general conformal Killing spinor $\psi^a(x)$ which grows at most linearly at infinity the Ward identities have the schematic form
\begin{equation}
\label{mainward}
\sum_i \psi(x_i)\langle {\cal O}_1(x_1)\ldots[{\bf Q},{\cal O}_i\}(x_i)\ldots{\cal O}_n(x_n)\rangle+\psi'(x_i)\langle {\cal O}_1(x_1)\ldots[{\bf S},{\cal O}_i\}(x_i)\ldots{\cal O}_n(x_n)\rangle=0
\end{equation}
Here we have not shown explicitly the spinor indices of $\psi$ and how they are contracted with the supercharges in order not to clutter the notation. Also by $[\ldots,\dots\}$ we mean commutator or anticommutator depending on whether the operator ${\cal O}$ is bosonic or fermionic. 

It is important to notice that we can always choose
$\psi^a(x)$ to vanish at some particular point $x_i$ and then the
corresponding term proportional to $[{\bf Q},{\cal O}\}(x_i)$ does not
contribute to the Ward identity. This observation is quite crucial and it is
a basic fact on which our proof is based. Notice also that if the operators ${\cal O}$ are superconformal primaries, we have $[{\bf S}, {\cal O}\}(x_{i}) = 0$ and the Ward identity becomes particularly simple.

Notice a possible confusing point in the expression above. The
commutators appearing in the Ward identity we are using are of the
form $[{\bf Q},{\cal O}](x)$, and not $[{\bf Q},{\cal O}(x)]$.
The first expression corresponds to first computing the commutator
between ${\bf Q}$ and ${\cal O}(0)$ and then using the translation operators
to translate the result at $x$, while the second expression is the
commutator between ${\bf Q}$ and the translated operator ${\cal O}(x)$.

These two expression can be different. For example, if ${\cal O}$ is a
superconformal primary, the commutator $[{\bf S},{\cal O} (x)]$ vanishes only
at the origin while $[{\bf S },{\cal O}](x)$ clearly vanishes everywhere.

\subsection{Chiral primary 3-point functions}

Now we come to the correlators, whose non-renormalization we aim to
prove.  These are 3-point functions of chiral primary operators
i.e. operators belonging to ``short'' multiplets of the superconformal
algebra. In theories with extended supersymmetry such operators must
fall into representations of the non-abelian R-symmetry. For example,
in the ${\cal N}=4$ SYM the R-symmetry is $SU(4)$ while in 2d CFTs
with ${\cal N}=(4,4)$ it is $SU(2)_L\otimes SU(2)_R$. Hence the chiral
primary operators are labeled by the representation ${\cal R}$ of the
R-symmetry and also by a set of additional indices $\vec{m}$ that
denote the specific element of the representation. As we will explain
later, the general structure of the 3-point function is
\begin{equation}
\label{generaltpf}
\langle \phi_I^{({\cal R}_1,\vec{m}_1)}(x_1)\,\,
 \phi_J^{({\cal R}_2,\vec{m}_2)}(x_2)\,\,
 \phi_K^{({\cal R}_3,\vec{m}_3)}(x_3)
\rangle = C_{IJK} \times \left({\rm group \,\, theoretic\,\, factors}\right)
\end{equation}
where the indices $I, J, K$ label various irreducible representations of the R-symmetry group. The only dynamical information is in the coefficients
$C_{IJK}$, which are precisely the coefficients whose independence of
the coupling we need to prove. The ``group theoretic factors'' above
contain both R-symmetry related factors, as well as the $x$-dependence
of the correlator which is completely fixed by conformal invariance.

Given the general form \eqref{generaltpf} of these 3-point functions
it becomes clear that we can isolate the desired coefficient $C_{IJK}$
by evaluating the correlator for specific alignments of the ${\vec
  m}$'s, as long as the corresponding group theoretic factor is
non-zero. In particular --- as we will explain in more detail later
--- it is possible to choose the operator at $x_2$ to be a ``highest
weight'' state in the representation ${\cal R}_2$ and the one at $x_3$
to be a ``lowest weight'' state in ${\cal R}_3$, while the one at $x_1$
will be ``mixed'' i.e. will have weight $\vec{m}$ which is neither
highest nor lowest. So we have that
$$
C_{IJK} \sim \langle \phi_I^{({\cal R}_1,\vec{m})}(x_1)\,\,
 \phi_J^{({\cal R}_2,+)}(x_2)\,\,
 \phi_K^{({\cal R}_3,-)}(x_3)
\rangle
$$ where $+, -$ denote the highest and lowest weight state
respectively.

The constant of proportionality depends on group theoretic factors and
is not relevant for us --- as long as it is non-zero. Also notice
that from the point of view of chiral primaries in ${\cal N}=1$
theories, the operator at $x_2$ would be ``chiral primary'', the one
at $x_3$ would be ``anti-chiral primary'' while the one at $x_1$ would
be neither chiral nor anti-chiral.

\subsection{Null vectors in short multiplets}

Before we proceed we need to make one more observation. The highest
weight state of a short representation is annihilated by some of the
supercharges. The lowest weight state is annihilated by the conjugate
supercharges. However, ``intermediate'' weight states in short
representations are generally not annihilated by {\it any} of the
supercharges.

While they are not annihilated by supercharges, these intermediate
states satisfy ``nullness conditions'', by which we mean that certain
linear combinations of superconformal descendants of intermediate
weight states in the multiplet are zero. These can be derived by
starting with the nullness conditions of the highest weight state
$[{\bf Q},\phi^{({\cal R},+)}\} = 0$ and acting on it with lowering
operators of the R-symmetry algebra. Using the Jacobi identity these
operators act both on the ${\bf Q}$ and on the chiral primary. Acting
with such lowering operators repeatedly we get conditions which have
the following general form \be
\label{nullcon}
[{\bf Q}\,,\, \phi^{({\cal R},\vec{m})}\}= \sum_i \,c_i\,[{\bf Q}'_i\,,\,\phi^{({\cal R},\vec{m}_i')}\}
\ee
where ${\bf Q}'_i$ are supercharges with R-symmetry weights different from those of ${\bf Q}$ and of course some of the $c_i$'s may be zero. The operators $\phi^{({\cal R},\vec{m}_i')}$ are in the same multiplet as $\phi^{({\cal R},\vec{m})}$ but have different R-symmetry weight. 

This condition will perhaps become more clear once we study it in specific theories.

\subsection{Supersymmetric marginal deformations}

The final element that we need is that the marginal deformations that
we are interested in are of special kind, they are deformations that
preserve not only conformal invariance but also
supersymmetry. Imposing that superconformal invariance is preserved
implies that the marginal operator must be a descendant of an
(anti)-chiral primary. Let us illustrate this with a few examples.

In 2d ${\cal N}=(2,2)$ theories, the supersymmetric marginal
deformations are of the form\footnote{Here the ${\bf G}$'s are the supercharges which will be defined in more detail section 4.}
$\{{\bf G}_{-\frac12}^- ,[\overline{\bf G}_{-\frac12}^-, \phi]\}$ and
$\{{\bf G}_{-\frac12}^+ ,[\overline{\bf G}_{-\frac12}^+ ,
    \overline{\phi}]\}$ where $\phi,\overline{\phi}$ are chiral
primaries in the $(c,c)$ and $(a,a)$ rings respectively, with
conformal dimension $(\frac12,\frac12)$, and also of the form
$\{{\bf G}_{-\frac12}^-,[ \overline{\bf G}_{-\frac12}^+ ,\psi]\}$ and
$\{{\bf G}_{-\frac12}^+, [\overline{\bf G}_{-\frac12}^- ,\overline{\psi}]\}$
  where $\psi,\overline{\psi}$ are chiral primaries in the $(a,c)$ and
  $(c,a)$ rings respectively, again with conformal dimension $(\frac12,\frac12)$.

Another example is the ${\cal N}=4$ SYM in 4d. There is only one
(complex) marginal coupling ${\cal O}_\tau$ preserving the full ${\cal
  N}=4$ supersymmetry, corresponding to changes of the complexified
gauge coupling $\tau = \frac{\theta}{2\pi} + i\frac{4\pi}{g^2}$. The
moduli space of this CFT is the upper half $\tau$ plane, modded out by
the appropriate S-duality group. The operator ${\cal O}_\tau$ is the (holomorphic part of the) Lagrangian
density. The important thing for us is that it can be written as
\be
\label{marginaln4}
{\cal O}_\tau = \{{\bf Q},[{\bf Q},\{{\bf Q},[{\bf Q},{\rm
      Tr}(Z^2)]\}]\} \ee where $Z$ is one of the complex adjoint
scalars. Here we did not write explicitly the indices of the
supercharges --- details can be found in appendix B.2. Notice that
these supercharges are all of the same chirality so they
(anti)-commute and their order is not important.

Instead of giving more examples, let us emphasize the
main point: supersymmetric marginal operators can be
written as
$$
{\cal O} = \{{\bf Q}, \Lambda\}
$$ where $\Lambda$ is {\it some} operator and ${\bf Q}$ is a
supercharge which annihilates either highest, or lowest weight
states. The operator $\Lambda$ is a descendant of chiral primaries of
specific conformal dimension (the details depend on the theory).

Finally, let us remind that the marginal operator has to be a singlet
of the R-symmetry of the theory. If not, it would break part of the supersymmetry.

\section{Outline of the proof}
\label{sec:outline}

Now we have collected all the ingredients and we can put them together
to give an outline of the proof. The (theory-specific) details will be
presented in the next sections.  \vskip5pt {\bf Step 1:} We isolate
the dynamical part of the 3-point function by aligning the chiral primaries so that one of
them is highest weight, the other lowest and the third
intermediate. So we have
$$
C_{IJK} \sim \langle  \phi_I^{({\cal R}_1,\vec{m})}(x_1)\,\,
 \phi_J^{({\cal R}_2,+)}(x_2)\,\,
 \phi_K^{({\cal R}_3,-)}(x_3)
\rangle
$$

\vskip5pt
{\bf Step 2:} We write the marginal operator corresponding to the change of a marginal coupling $g$ as ${\cal O} = \{{\bf Q}, \Lambda\}$. Hence  we would like to prove the vanishing of
$$
\nabla_g C_{IJK}  \sim  \int d^dx\, \langle \{{\bf Q}, \Lambda\}(x)\,\, \phi_I^{({\cal R}_1,\vec{m})}(x_1)\,\,
 \phi_J^{({\cal R}_2,+)}(x_2)\,\,
 \phi_K^{({\cal R}_3,-)}(x_3)
\rangle
$$
Let us denote the {\it integrand} by ${\cal I}$, on which we now focus.
\vskip5pt
{\bf Step 3:} Without loss of generality we can assume that ${\bf Q}$ annihilates the highest weight operator at $x_2$. Then we use the superconformal Ward identity 
\eqref{mainward} with a spinor $\psi^a(x)$ vanishing at $x_3$ to move ${\bf Q}$ away from the point $x$ and we get that\footnote{Again, by $[\ldots,\ldots\}$ we mean the commutator (or anticommutator) if the operator is bosonic (or fermionic).}
$$
{\cal I} \sim \langle  \Lambda(x)\,\, [{\bf Q}, \phi_I^{({\cal R}_1,\vec{m})}\}(x_1)\,\,
 \phi_J^{({\cal R}_2,+)}(x_2)\,\,
 \phi_K^{({\cal R}_3,-)}(x_3)
\rangle
$$
The important point here is that there is no other contribution to the Ward identity\footnote{Notice that $\phi_I,\phi_J,\phi_K$ are all superconformal primaries, so they are annihilated by the ${\bf S}$'s.}.
\vskip5pt
{\bf Step 4:} We use the ``nullness condition'' \eqref{nullcon} for the operator at $x_1$ to rewrite this as
$$
{\cal I} \sim \sum_{'}\langle \Lambda(x)\,\, [{\bf Q}',\phi_I^{({\cal R}_1,\vec{m'})}\}(x_1)\,\,
 \phi_J^{({\cal R}_2,+)}(x_2)\,\,
 \phi_K^{({\cal R}_3,-)}(x_3)
\rangle
$$
where $\vec{m'}$ is some other element of the same representation and ${\bf Q}'$ supercharges with R-symmetry weight different from those of ${\bf Q}$.
\vskip5pt
{\bf Step 5:} The set of supercharges ${\cal A}$ can be partitioned into two disjoint sets ${\cal A} = {\cal A}_{+} \cup {\cal A}_{-}$, where the charges in $ {\cal A}_{+}$ annihilate the highest weight states and the charges in ${\cal A}_{-}$ annihilate the lowest weight states. If ${\bf Q}' \in  {\cal A}_{+}$ then we use the Ward identity with a spinor vanishing at $x_3$ to move ${\bf Q'}$ away from $x_1$. If ${\bf Q}' \in  {\cal A}_{-}$ then we choose a spinor which vanishes at $x_2$. In both cases we have
$$
{\cal I} \sim \sum_{'}\langle  \{{\bf Q}',\Lambda\}(x)\,\, \phi_I^{({\cal R}_1,\vec{m'})}(x_1)\,\,
 \phi_J^{({\cal R}_2,+)}(x_2)\,\,
 \phi_K^{({\cal R}_3,-)}(x_3)
\rangle
$$

\vskip5pt
{\bf Step 6:} Remarkably the quantum numbers conspire in such a way that in the theories that we study $\{{\bf Q}', \Lambda\} = 0$. Hence
$$
{\cal I} =0\qquad \Rightarrow \qquad \nabla_g\, C_{IJK} = 0
$$
This completes the proof.

Here we have skipped many details which will be presented in the next
sections, since they are theory-dependent.

\section{Two-dimensional CFTs with ${\cal N}=(4,4)$ supersymmetry}
\label{sec:2d}

In this section we present the non-renormalization theorem for 3-point
functions of chiral primaries in two-dimensional ${\cal N}=(4,4)$
superconformal field theories, generalizing and completing the results
of \cite{arXiv:0809.0507}.

In the first subsection we describe the short multiplets in these
theories and review the general form of the 3-point function of chiral
primaries. In the second subsection we prove the non-renormalization
theorem.

\subsection{Short representations and their 3-point functions}

The R-symmetry of the ${\cal N}=(4,4)$ superconformal algebra is
$SU(2)_L \otimes SU(2)_R$. The left moving supercharges are denoted by
${\bf G}_{-\frac12}^{a r}$. Here the index $a=\pm$ denotes the
$J^3$ eigenvalue with respect to the left-moving $SU(2)_L$ R-symmetry,
while the index $r=\pm$ denotes the eigenvalue of the supercharge
under a left $SU(2)$ outer automorphism of the ${\cal N}=4$
algebra. The right-moving supercharges have similar structure. We
refer the reader to \cite{arXiv:0809.0507} for more details.

Representations of the algebra are labeled by the conformal dimension
$\{h,\overline{h}\}$ and the R-symmetry representation
$\{j,\overline{j}\}$ of the superconformal primaries\footnote{i.e.
  operators annihilated by all ${\bf G}^{ab}_{n},\,\,n>0$.} of the
multiplet. Notice that a given multiplet contains several
superconformal primaries which differ by their $SU(2)_L\otimes
SU(2)_R$ quantum numbers. Unitarity requires
$$
h\geq j,\qquad \overline{h}\geq \overline{j}
$$ Multiplets which saturate the bound are ``short" and are usually
called ``chiral primary" multiplets.

To simplify notation, in the following we will sometimes write only the quantum numbers of the left-moving sector. For a multiplet characterized
by conformal dimension $h$ and R-symmetry quantum number $j$, we have the following set of superconformal primaries
$$
\phi^{(j,m)},\qquad m=-j,\ldots,+j
$$ which differ by their $J^3$ eigenvalue $m$. All these operators are
superconformal primaries, they have conformal dimension $h$ and can be
recovered from the ``highest weight" state of the multiplet by
acting with $SU(2)_L$ lowering operators
$$
\phi^{(j,m)} \sim \overbrace{ [J^-\,,\ldots[J^-\,}^{j-m},\,\phi^{(j,j)}]\ldots]
$$

The ``highest weight" operator of a short multiplet $\phi^{(j,j)}$ is
annihilated by some of the supercharges
$$
[{\bf G}^{+r}_{-\frac12} \,\,,\,\, \phi^{(j,j)}\} = 0,\qquad r = +,-
$$
and similarly for the ``lowest weight" one $\phi^{(j,-j)}$
$$
[{\bf G}^{-r}_{-\frac12} \,\,,\,\, \phi^{(j,-j)}\} = 0,\qquad r = +,-
$$ 
The other members of the short multiplet $\phi^{(j,m)}$ with $m
\neq \pm j$ {\it are not} annihilated by any of the left moving
supercharges. They do however satisfy nullness conditions, which can be
derived by starting with $[{\bf G}_{-\frac12}^{+r}\,\,,\,\, \phi_I^{(j,j)}\} = 0$ and acting with lowering operators $J^-$. This
leads to the following relation\footnote{This equation is proven in appendix \ref{sec:nullconditions}.}
\begin{equation}
\label{descen}
[{\bf G}^{+r}_{-\frac12}\,\,,\,\, \phi_I^{(j,n)} \}\quad \sim \quad [{\bf
    G}^{-r}_{-\frac12} \,\,,\,\, \phi_I^{(j,n+1)}\}
\end{equation}
where the constant of proportionality is nonzero as long as $n<j$.

Notice that here there is some potentially confusing terminology: from
the ${\cal N}=(4,4)$ point of view, all the operators $\phi^{(j,m)}$ are
sometimes called ``chiral primaries", since they all belong to the same
short multiplet. If however we consider an ${\cal N}=(2,2)$ subalgebra
then the operator $\phi^{(j,j)}$ would be called ``chiral", the
operator $\phi^{(j,-j)}$ ``antichiral" and the other operators
$\phi^{(j,m)}$ with $m\neq \pm j$ would be neither chiral nor
antichiral.

Let us write the general form of the 3-point function of chiral
primary operators. We have
\begin{equation}
\label{mainthree}
\begin{split}
& \langle \phi_I(x_1)\,\, \phi_J(x_2)\,\,\phi_K(x_3)\rangle = C_{IJK}  \left(\begin{array}{ccc}j_1  &
j_2 & j_3\\
 m_1 &m_2 & m_3\end{array}\right)
  \left(\begin{array}{ccc} \overline{j}_1  &
\overline{j}_2 & \overline{j}_3
  \\ \overline{m}_1 &\overline{m}_2 & \overline{m}_3\end{array}\right)\cr &
\times \frac{1}{x_{12}^{(j_1+j_2-j_3)} x_{23}^{(j_2+j_3-j_1)} x_{13}^{(j_1+j_3-j_2)}}
\frac{1}{\overline{x}_{12}^{(\overline{j}_1+\overline{j}_2-\overline{j}_3)} \overline{x}_{23}^{(\overline{j}_2+
\overline{j}_3-\overline{j}_1)} \overline{x}_{13}^{(\overline{j}_1+\overline{j}_3-\overline{j}_2)}}
\end{split}
\end{equation}
Here we did not write explicitly the $SU(2)_L\otimes SU(2)_R$ quantum numbers on the LHS of the equation.

The $x$-dependence in \eqref{mainthree} is fixed by conformal
invariance in terms of the conformal dimension of the operators. The
dependence on the quantum numbers $(j,m\,; \overline{j},\overline{m})$
is fixed by the $SU(2)_L\otimes SU(2)_R$ R-symmetry and is expressed
by the 3-j symbols presented above. All the dynamical information is
encoded in the coefficient $C_{IJK}$, which as we can see only depends
on the choice of chiral primary representations $I,J,K$ and not on the
specific representatives from each of them (i.e. does not depend on
the R-symmetry quantum numbers $m,\overline{m}$).

Our goal is to show that the constants $C_{IJK}$ do not depend on the
coupling constants of the CFT.

Going back to the distinction between a ``chiral primary" in ${\cal
  N}=(2,2)$ theories and a ``chiral primary" in ${\cal N}=(4,4) $ theories let
us notice the following: in ${\cal N}=(2,2)$ theories R-charge
conservation requires that the three operators satisfy the condition
$j_3=j_1+j_2$ (or permutations) - and similarly for the right-moving
sector.  These would be ``extremal" 3-point functions of chiral
primaries from the ${\cal N}=(4,4)$ point of view. However in ${\cal N}=(4,4)$
theories there are also 3-point functions of chiral primaries which
are not in the extremal case.

In \cite{arXiv:0809.0507} a non-renormalization theorem for 3-point functions was
proven for the special case where the three chiral primary multiplets
satisfy $j_3=j_1+j_2$ (or permutations) i.e. for the ``extremal
case". In that case the 3-point function can be viewed as a 3-point
function of chiral primaries of an ${\cal N}=(2,2)$ subalgebra. In the
more general case where $j_3 \neq j_1+j_2$ this is not possible. There
is no way to align all three operators so that they are all in the
chiral ring of a given ${\cal N}=(2,2)$ subalgebra. Nevertheless, the
``non-extremal" 3-point functions also seem
to be protected and thus should obey some non-renormalization theorem, which we will prove in the next subsection.

\subsection{The non-renormalization theorem in 2d}
\medskip
Theories with ${\cal N}=(4,4)$ supersymmetry in two dimensions have a moduli space of marginal deformations which is locally of the form $\frac{SO(n,4)}{SO(n)\times SO(4)}$ \cite{Seiberg:1988pf}. Here $n$ is the number of chiral primary multiplets which transform in the $(1/2,1/2)$ representation of the $SU(2)_L\otimes SU(2)_R$ R-symmetry group. 

Let us consider the 3-point function of operators which belong to
chiral primary multiplets
$$
 \langle \phi_I (x_1)\,\, \phi_J (x_2)\,\,\phi_K(x_3)\rangle
$$
\smallskip
\noindent where for simplicity we do not write any R-symmetry indices,
this is supposed to be a condensed notation for \eqref{mainthree}. Let
us also consider a marginal operator ${\cal O}$ corresponding to the
change of a marginal coupling constant $g$. By definition we have
$$
\nabla_g\, \langle \phi_I (x_1)\,\, \phi_J
(x_2)\,\,\phi_K(x_3)\rangle \equiv \int d^2x \,\langle {\cal
  O}(x)\,\, \phi_I (x_1)\,\, \phi_J(x_2) \,\,\phi_K(x_3)\rangle
$$
In order to prove that the 3-point functions are independent of the coupling we have to show that the expression
above vanishes. We will actually prove a stronger statement, namely that
$$
{\cal I} \equiv \langle {\cal O}(x)\,\, \phi_I(x_1)\,\, \phi_J(x_2) \,\,\phi_K(x_3)\rangle = 0
$$
even without integrating over $x$. We will follow the steps outlined in section \ref{sec:outline}.
\smallskip
In order to prove this we will use two properties of the ${\cal N}=(4,4)$ algebra
\vskip10pt

First, we  exploit the $SU(2)_L\otimes SU(2)_R$ structure of the correlator \eqref{mainthree}. If we simply want
to compute the 3-point function $C_{IJK}$ --- or rather to prove that it is independent of the coupling --- we are free to evaluate
the correlator for any alignment of the operators for which the 3j symbols are non-vanishing. Hence we will choose the representatives of
the other chiral primaries in the following way
$$
{\cal I}\sim  \langle {\cal O}(x)\,\, \phi_I^{(j_1,n)}(x_1)\,\, \phi^{(j_2,j_2)}_J(x_2) \,\,\phi^{(j_3,-j_3)}_K(x_3)\rangle
$$
where $n=j_3-j_2$. The constant of proportionality is some (non-vanishing) group-theoretic factor which is of no
interest for our argument. Notice that from the point of view of an ${\cal N}=2$ subalgebra the operator at $x_2$ is
``chiral primary", the operator at $x_3$ is ``anti-chiral primary" while the operator at $x_1$ is neither chiral on
antichiral.

Second,  without loss of generality\footnote{This is a general property of ${\cal N}=(4,4)$ SCFTs which was discussed in detail
in \cite{arXiv:0809.0507}.} we can assume that the marginal operator can be written as ${\cal O} = \{{\bf G}_{-\frac12}^{+r}\,,\,
[{\bf \overline{G}}_{-\frac12}^{\,+s} \,,\, \overline{\phi}]\}$
where $\overline{\phi}$ is an element of a chiral primary multiplet of conformal weight $\left(\frac12,\frac12\right)$ and which is
aligned to have $(J^3,\overline{J}^{\,3}) = \left(-\frac12,-\frac12\right)$.

Then we have that
$$
{\cal I} \sim \langle
\left(\{{\bf G}_{-\frac12}^{+r}\,,\,
[{\bf \overline{G}}_{-\frac12}^{\,+s} \,,\, \overline{\phi}]\} \right)(x) \,\,
\phi^{(j_1,n)}_I(x_1)
\,\,\phi^{(j_2,j_2)}_J(x_2)\,\,\phi^{(j_3,-j_3)}_K(x_3)\rangle
$$
Using a superconformal Ward identity \eqref{mainward} for ${\bf G}_{-\frac12}^{+r}$ with
a conformal Killing spinor vanishing at the point $x_3$ we find that this can be written as
$$
{\cal I} \sim \langle [{\bf \overline{G}}_{-\frac12}^{+s}\,,\, \overline{\phi}](x) \,\,[{\bf G}_{-\frac12}^{+r}\,,\,  \phi^{(j_1,n)}_I\}(x_1)
\,\,\phi^{(j_2,j_2)}_J(x_2)\,\,\phi^{(j_3,-j_3)}_K(x_3)\rangle
$$
where the constant of proportionality in this expression is different from zero. Here we used that ${\bf G}_{-\frac12}^{+r}$
annihilates the operator at $x_2$.

Now we use the nullness condition \eqref{descen} to rewrite it as
$$
{\cal I} \sim \langle [{\bf \overline{G}}_{-\frac12}^{+s}\,,\, \overline{\phi}](x)\,\,
[{\bf G}_{-\frac12}^{-r}\,,\, \phi_I^{(j_1,n+1)}\}(x_1)\,\,
\phi_J^{(j_2,j_2)}(x_2) \,\,\phi_K^{(j_3,-j_3)}(x_3)\rangle
$$
Finally we use a superconformal Ward identity for ${\bf G}_{-\frac12}^{-r}$ with a conformal Killing spinor which vanishes at the point $x_2$. All
other operators do not contribute because they are annihilated by ${\bf G}_{-\frac12}^{-r}$, hence we find
$$
{\cal I} = 0
$$
This proves that 3-point functions of chiral primaries are independent of the coupling constant.

Notice that it would
not be possible to apply a similar argument to prove non-renormalization of 4- and higher point functions of
chiral primaries (unless they are extremal \cite{arXiv:0809.0507}, see also \cite{Cardona:2010qf}, \cite{Kirsch:2011na})
--- which is of course consistent, since we know that such correlators {\it do} depend on the coupling constants.

\section{Four-dimensional ${\cal N}=4$ SCFTs}

The same type of argument can be used to prove the non-renormalization
of 3-point functions of 1/2 BPS chiral primaries in four-dimensional
SCFTs with ${\cal N}=4$ supersymmetry.

\subsection{Short representations}

Now the R-symmetry is $SU(4)$. We choose a basis for its Cartan
subalgebra. The short representations that we are interested in are
those with Dynkin labels $[0,k,0]$, Lorentz spin $(j,\overline{j}) =
(0,0)$ and conformal dimension $\Delta = k$. These are the ``1/2 BPS'' operators
of the ${\cal N}=4$ algebra. In terms of Young
tableaux for $SU(4)$ these representations correspond to tableaux with
$k$ columns of length 2 (we refer to appendix \ref{sec:oko} for more details). As before we denote the superconformal
primaries of such a multiplet by
$$
\phi^{(k, \vec{m})}
$$ where now $\vec{m}$ labels the weight of the state inside the
$SU(4)$ multiplet (i.e. $\vec{m}$ are the eigenvalues of the state
under the Cartan generators).  Of special interest will be the highest
and lowest weight states of any given representation, which we call
$\phi^{(k,\pm)}$. For example, in some conventions highest weight operators are ${\rm
  Tr}(Z^k)$ and their multi-trace products.

Let us recall some group theory (more details are given in appendix \ref{sec:groupt} and \ref{sec:su4}). We denote by $E_i$ the generators of
$SU(4)$ corresponding to positive simple roots, or raising operators. The highest weight
state satisfies $[E_i\,,\, \phi^{(k,+)}] = 0$. Other operators in the same
$SU(4)$ multiplet can be recovered starting from $\phi^{(k,+)}$ and
acting with the lowering operators $E_i^{\dagger}$
$$
\phi^{(k,\vec{m})} \sim  [E_{i_n}^\dagger\,,\ldots[ E_{i_1}^\dagger\,,\, \phi^{(k,+)}]\ldots]
$$ where the product is some specific combination of the ``negative
simple roots", perhaps with repeated appearances.

Of course equivalently we can start from the lowest weight state and
get the same state by acting with ``raising" operators.
$$
\phi^{(k,\vec{m})} \sim  [E_{i_n}\,,\ldots[ E_{i_1}\,,\, \phi^{(k,-)}]\ldots]
$$

It is a group-theoretic fact that in a tensor product of the form
$[0,k_1,0]\otimes [0,k_2,0]$ any representation of the form $[0,k_3,0]$
appears either one time or none\footnote{If $k_1,k_2,k_3$ satisfy the
  triangle (in)-equality and $\frac{k_1+k_2+k_3}{2}$ is an integer,
  then the representation appears one time. Otherwise it does not
  appear.}. Hence the general form of a 3-point function is
$$
\langle \phi_I^{(k_1,\vec{m}_1)}(x_1)\,\,\phi_J^{(k_2,\vec{m}_2)}(x_2)\,\,\phi_K^{(k_3,\vec{m}_3)}(x_3)\rangle
=C_{IJK} {\bf G}(k_1,\vec{m}_1; k_2,\vec{m}_2; k_3,\vec{m}_3)$$
$$\times \frac{1}{|x_{12}|^{k_1+k_2-k_3}|x_{23}|^{k_2+k_3-k_1}|x_{13}|^{k_1+k_3-k_2}}
$$ where ${\bf G}(k_1,\vec{m}_1; k_2,\vec{m}_2; k_3,\vec{m}_3)$ is the (unique) 
$SU(4)$ Clebsh--Gordan coefficient for three representations of the
type $[0,k,0]$ i.e. a group-theoretic factor.  The dynamical information is
encoded in the coefficient $C_{IJK}$.

Notice that, as emphasized previously in the paper, it is only the highest and lowest weight states of the short multiplets that are annihilated by supercharges. ``Intermediate weight'' states are generally not annihilated by any of the supercharges (though they lead to certain ``nullness conditions'' as explained earlier). For example, while the superconformal primary operators of the form
\be
C_{i_{1}\ldots i_{n}} \mathrm{Tr}(\phi^{i_{1}} \cdots \phi^{i_{k}})
\ee
with $C$ symmetric and traceless are members of 1/2 BPS multiplets, for generic choice of such symmetric traceless $C$, they are {\it not} annihilated by any supercharges. Only if $C$ is chosen so that the corresponding operator is highest or lowest weight state with respect to $SU(4)_R$ is the operator annihilated by 1/2 of the supercharges\footnote{In \cite{Basu:2004nt} it was incorrectly assumed that all superconformal primaries of the 1/2 BPS multiplet are annihilated by half of the supercharges, hence the proposed proof of the non-renormalization theorem in \cite{Basu:2004nt} is incomplete.}.

\subsection{The non-renormalization theorem in 4d}

First let us choose a basis of the left chiral supercharges so that
they have definite weight under the Cartan subalgebra\footnote{And also a definite weight under the $J_3$ of the $SU(2)_L$ part of the Lorentz group.}. We denote these
left chiral supercharges as ${\bf Q}^i_a$ where the index
$i=1,\ldots 4$ is the $SU(4)$ and $a$ the Lorentz index.

The theory has an exactly marginal operator ${\cal O}_\tau$
corresponding to the change of the complexified coupling constant
$\tau = \frac{\theta}{2\pi} + i \frac{4\pi}{g^2}$. As mentioned before and explained in detail in appendix \ref{sec:oto} --- this operator can be written as
\begin{equation}
\label{marginal}
 {\cal O}_\tau = ({\bf Q})^4 \phi^{(2,+)} 
\end{equation} 
where only four of the left-chiral supercharges act on the highest
weight state. The notation $({\bf Q})^4$ means the nested
(anti)-commutator, as in equation \eqref{marginaln4}, we hope this is obvious. Notice that the left
chiral supercharges anticommute among themselves so we do not need to
worry about the order with which they act on an operator.

The set of left chiral supercharges ${\cal A}$ can be partitioned into two disjoint sets ${\cal A} = {\cal A}_{+} \cup {\cal A}_{-}$, where the charges in $ {\cal A}_{+}$ annihilate the highest weight states of the 1/2 BPS multiplets and the charges in ${\cal A}_{-}$ annihilate the lowest weight states. The set of supercharges which appear in \eqref{marginal} is simply ${\cal A}_{-}$, and any other left chiral supercharge in ${\cal A}_{+}$ annihilates
the operator $\phi^{(2,+)}$. This will be important below.

Consider now the change of a 3-point function under a deformation by
${\cal O}_\tau$. We will show that
$$
{\cal I} \equiv \langle {\cal O}_\tau(x)\,\,\phi_I(x_1)\,\, \phi_J(x_2)\,\, \phi_K(x_3)\rangle =0
$$

As before we can choose the $SU(4)$ alignment of the operators in such a way that
\begin{equation}
\label{alignfour}
{\cal I} \sim \langle {\cal O}_\tau(x)\,\,\phi_I^{(k_1,\vec{m}_1)}(x_1)\,\,\phi_J^{(k_2,+)}(x_2)\,\,\phi_K^{(k_3,-)}(x_3)\rangle
\end{equation}
where the operator at $x_2$ is a highest weight state, the one at
$x_3$ is lowest weight and the one at $x_1$ is of some general weight
in the representation $k_1$. Using the form of the marginal operator
we have
$$
{\cal I} \sim \langle \left( ({\bf Q})^4 \phi^{(2,+)}\right)(x)\,\,\phi_I^{(k_1,\vec{m}_1)}(x_1)\,\,\phi_J^{(k_2,+)}(x_2)
\,\,\phi_K^{(k_3,-)}(x_3)\rangle
$$ Notice that the four supercharges acting on the operator at $x$ are
all left chiral so they (anti)-commute and their order is not
important. As we mentioned above we call this set of supercharges ${\cal A}_{-}$. Also notice that all of these four supercharges annihilate
the operator at $x_3$.

We take one of them, let us call it ${\bf Q}^\star$ and move it away
using the Ward identity. We choose the conformal Killing spinor to
vanish at the point $x_2$. Hence the correlator becomes
$$
{\cal I} \sim \langle \left( ({\bf Q})^3 \phi^{(2,+)}\right)(x)\,\,\left([{\bf Q}^\star,\phi_I^{(k_1,m_1)}]\right)(x_1)\,\,
\phi_J^{(k_2,+)}(x_2)\,\,\phi_K^{(k_3,-)}(x_3)\rangle
$$

Now we will use the analogue of \eqref{descen} coming from the fact that ${\bf Q}^\star$ annihilates the lowest weight
state of the representation $k_1$, that is.
\begin{equation}
\label{nullstate}
[{\bf Q^\star},\phi_I^{(k_1,\vec{m}_1)}] = \sum_{j\neq\star} [{\bf Q}^j, {\cal X}_j]
\end{equation}
where all supercharges in the sum on the RHS are left chiral and different from ${\bf
  Q}^\star$ and ${\cal X}_j$ is either one of the elements of the
multiplet $\phi_I^{(k_1,\vec{m}_j)}$ or perhaps zero\footnote{In either
  case the operator ${\cal X}$ is annihilated by the
  $\overline{S}$'s.}. This important relation is proven in 
appendix \ref{sec:nullconditions}.

Next, for each of these ${\bf Q}^j$'s we apply the Ward identity
\eqref{mainward} again. There are two possibilities: \vskip5pt 1)
${\bf Q}^j$ is in ${\cal A}_{-}$: in this case we use \eqref{mainward} with a
spinor vanishing at $x_2$. We do not get any contribution from $x_3$
because the operator is annihilated by the supercharges in ${\cal
  A}_{-}$. We do not get any contribution from $x$ because the supercharge
is already there, so it squares to zero.  \vskip5pt 2) ${\bf Q}^j$ is 
in ${\cal A}_{+}$: then this supercharge annihilates operators of the form
$\phi^{(k,+)}$. Then we use \eqref{mainward} superconformal Ward identity with a
spinor vanishing at $x_3$ and we get zero.  \vskip5pt So in all cases
the contribution is zero. Hence
$$
{\cal I}=0 \qquad \Rightarrow \qquad \nabla_\tau C_{IJK} =0
$$ Exactly the same argument can be applied for the marginal operator
$\overline{\cal O}_\tau\equiv(\overline{\bf Q})^4 \phi^{(2,-)}$. So all in all
the 3-point functions are not renormalized and this completes the proof.

 Notice that this argument fails --- as expected --- if we try to prove the non-renormalization of $n$-point functions of chiral primaries with $n>3$ (unless they are "extremal"). The last step of the proof relied on the fact that there was at most one operator which was not annihilated by the supercharge involved in the Ward identity. We chose the Killing spinor to vanish at the point where this operator was inserted. If there were more than one operators not annihilated by the supercharge, it would not be possible to simultaneously "hide" their contributions to the Ward identity by choosing the Killing spinor appropriately.

\section{Extremal correlators}

Similar arguments can be used to show that a certain class of higher $n$-point functions are not renormalized. These are the so-called ``extremal correlators'' i.e. correlators where all chiral primaries are aligned to be ``highest weight'' except for one that is aligned to be ``lowest weight'' and which ensures R-charge neutrality
$$
\langle \phi_1^{({\cal R}_1,+)}(x_1)\,\,\phi_2^{({\cal R}_2,+)}(x_2)\ldots\phi_n^{({\cal R}_n,-)}(x_n)
\rangle
$$
Charge conservation shows that the operators must satisfy $\Delta_n = \sum_{i=1}^{n-1}
\Delta_i$. 

That such correlators are not renormalized in 2d ${\cal N}=(4,4)$ theories was proven in \cite{arXiv:0809.0507}. The proof was based on the observation that in these theories a marginal operator can always be written as ${\cal O} = [{\bf G}^{-r}_{-\frac12},\Lambda
]$. Then we can consider 
\begin{equation}
\label{blabla}
\langle {\cal O}(z) \,\,\phi_1^{({\cal R}_1,+)}(x_1)\,\,\phi_2^{({\cal R}_2,+)}(x_2)\ldots\phi_n^{({\cal R}_n,-)}(x_n)
\rangle
\end{equation}
and use a Ward identity with a spinor vanishing at $x_n$ to move the supercharges away from $z$. The operators at $x_1,\ldots,x_{n-1}$ do not contribute since they are annihilated by $\overline{G}^{-r}_{-\frac12}$ and the operator at $x_n$ does not contribute because of the choice of the spinor in the Ward identity. Hence this correlator vanishes and the desired result is proven.

Let us quickly repeat the similar statement in ${\cal N}=4$ SYM. In that theory we have two marginal operators --- corresponding to changes of the coupling constant $g$
and the $\theta$-angle --- which can be combined into the holomorphic and anti-holomorphic operators  ${\cal O}_\tau, \overline{\cal O}_\tau$. One of these operators can be written as 
\begin{equation}
\label{antihol}
\overline{\cal O}_\tau = (\overline{\bf Q})^4 {\rm Tr}(\overline{Z}^2)
\end{equation}
where the supercharges $\overline{\bf Q}$ annihilate highest weight states of $SU(4)$. Hence we can use the Ward identity with a spinor vanishing at $x_n$ to show that the analogue of \eqref{blabla} in ${\cal N}=4$ vanishes. To complete the proof of the non-renormalization we also need to show that the same correlator vanishes for the marginal operator ${\cal O}_\tau$. We can use the fact that in ${\cal N}=4$ theories this marginal operator can also be written as
\begin{equation}
\label{hol}
{\cal O}_\tau = ({\bf Q})^4 {\rm Tr}(\overline{Z}^2)
\end{equation}
where the ${\bf Q}$'s are supercharges of left chirality. This may look confusing when compared to \eqref{antihol} and against our intuition from theories with less supersymmetry, but it is indeed a true statement (explained in appendix \ref{sec:oto})\footnote{Notice that the four ${\bf Q}$'s in \eqref{hol} {\it are not} the complex conjugates of the supercharges in \eqref{antihol}.}. The four supercharges in \eqref{hol} annihilate the highest weight states of $SU(4)$ of the form $\phi^{(k,+)}$. Hence the Ward identities can be used as above to show that the correlator vanishes.

All in all we have proved that extremal $n$-point functions of 1/2 BPS chiral primaries in four-dimensional ${\cal N}=4$ SCFTs are not renormalized.

\section{Other extensions}

In this section we list some immediate generalizations of our results. 

\subsection{Half-chiral states in 2d ${\cal N}=(4,4)$}

Interestingly, the argument in section \ref{sec:2d} relied only on one sector --- say the left moving one --- of the CFT. This implies that the same argument goes through without changes when applied to 3-point functions of operators that are in short multiplets of the left-moving $SU(2)_L$ and long multiplets on the right-moving one. Such operators are of the form (chiral, anything). Our argument shows that their 3-point functions are not renormalized as a function of the coupling constants. Notice that these states are related by spectral flow to states of the form (Ramond ground state, anything) which are precisely the microstates of the Strominger-Vafa black hole \cite{Strominger:1996sh}. It would be interesting to explore the possible applications of this statement.

Notice however that our arguments show that the 3-point functions of such states do not renormalize as a function of the coupling {\it assuming} that they remain chiral primaries during the deformation (i.e. that short multiplets do not combine and lift from the BPS bound). We have not addressed the issue of whether BPS states lift or not under marginal deformations.

\subsection{3-point functions in 2d ${\cal N}=(0,4)$ SCFTs}

Another interesting case is that of two-dimensional CFTs with $(0,4)$ supersymmetry. In string theory they arise on the worldvolume of bound states of M2/M5 branes wrapped on Calabi-Yau compactifications of M-theory and are relevant for the computation of the entropy of certain supersymmetric black holes \cite{Maldacena:1997de, Minasian:1999qn}. 

Theories with ${\cal N}=(0,4)$ supersymmetry are not very well understood, but it is clear that on their ``supersymmetric side'' they have operators in short representations, which are the analogue of the (anything, chiral) operators in $(4,4)$ CFTs. Our claim is that 3-point functions of such operators are not renormalized as a function of the coupling constants. This follows immediately from our proof, if we also remember that marginal operators in these theories can be written as ${\cal O} = 
[\overline{G}_{-\frac12}^{\pm r} ,\phi]$ and its conjugate, where $\phi$ is ``chiral primary'' with respect to the right moving supersymmetric side. Also, notice that the statement holds only for operators which do not lift from the BPS bound as we vary the coupling.  

\subsection{Less supersymmetric multiplets in 4d}

It would be interesting to generalize our results to 1/4 and 1/8 BPS operators in four dimensional $\mathcal{N} = 4$ SCFT. Unfortunately, the group theory structure of the correlators is much more intricate in this case. For example, the product of three 1/4 BPS scalar operators, which sit in $[q,p,q]$ representations of the $SU(4)$ R-symmetry group, contains many trivial representations. As an example, the product of three $[1,2,1]$ representations contains 5 distinct trivial representations. This means that the corresponding 3-point functions are not determined by a single numerical coefficient, unlike what happened in the 1/2 BPS case.

As a consequence, the first step of choosing an alignment cannot  be carried out in general. It is interesting to explore whether the rest of the proof extends at least for specific alignments. So let us consider a general 3-point function, aligned in a convenient way, and let us try to derive some necessary conditions for our proof to hold. It is clear that the highest-weight of such operators should be annihilated by at least one supercharge, so let us consider the product of three 1/8 BPS operators, so that the change of their 3-point function generated by ${\cal O}_{\tau}$ reads
\be
{\cal I} \sim \langle \left( ({\bf Q})^4 \phi^{(2,+)}\right)(x)\,\,\phi_I^{(\mathcal{R}_{1},\vec{m}_1)}(x_1)\,\,\phi_J^{(\mathcal{R}_{2},+)}(x_2)
\,\,\phi_K^{(\mathcal{R}_{3},-)}(x_3)\rangle
\ee
The charges appearing in $({\bf Q})^4$ are either ${\bf Q}^{3}$ or ${\bf Q}^{4}$. We can take one of the two\footnote{Remember that the supercharges ${\bf Q}^{i}$ are spinors, so they also carry a Lorentz index.} ${\bf Q}^{4}$'s (which annihilates the operator at $x_{3}$, since it is a lowest-weight) and move it using a Ward identity with a conformal Killing spinor that vanishes at $x_{2}$:

\be
{\cal I} \sim \langle \left( ({\bf Q})^3 \phi^{(2,+)}\right)(x)\,\,[{\bf Q}^{4}, \phi_I^{(\mathcal{R}_{1},\vec{m}_1)}](x_1)\,\,\phi_J^{(\mathcal{R}_{2},+)}(x_2)
\,\,\phi_K^{(\mathcal{R}_{3},-)}(x_3)\rangle
\ee
The null condition applied to the operator at $x_{1}$ will generically give supercharges ${\bf Q}^{i}$ with $i=1,2,3$, therefore if we want to use a Ward identity to argue that ${\cal I}$ vanishes, the operator at $x_{2}$ and $x_{3}$ should be 1/2 BPS operators\footnote{If the highest-weight is annihilated by ${\bf Q}^{i}$ with $i=1,2$, a simple argument based on unitarity bounds \cite{Dolan:2002zh} shows that it must be annihilated by $\overline{\bf Q}_{i}$ with $i=3,4$ as well.}. As a consequence, the proof seems to work only for the case $1/8\otimes 1/2\otimes 1/2$.

A simple application of the Berenstein--Zelevinsky triangles shows that a product of the form  $[p,q,r] \otimes [0,k_{1},0] \otimes [0,k_{2},0]$ contains the trivial representation only if $p=r$, which implies that the the operator at $x_{1}$ must be 1/4 BPS. In this case, if the trivial representation does appear, it appears only one time and the relative Clebsh--Gordan coefficient is unique. Furthermore, since the highest-weight of a 1/4 BPS operator is also annihilated by a right chiral supercharge $\overline{\bf Q}_{4}$, the proof works for the $\overline{\cal O}_{\tau}$ operator as well.

Summarizing, we were able to generalize the non-renormalization proof to the 3-point function of one 1/4 BPS operator and two 1/2 BPS operators, but the proof seems to fail in more general cases.

\section{Discussions}

We proved the non-renormalization of certain correlation functions of
chiral primary operators in 4d ${\cal N}=4$ and 2d ${\cal N}=(4,4)$
superconformal field theories. Our proof was based on the
superconformal Ward identities and not on superspace arguments. While
equivalent to the latter, we find that the direct proof offers some conceptual 
advantages.

It would be interesting to explore further more general correlators,
for example three point functions of 1/4 BPS operators, and see
whether an argument for their non-renormalization can be found. Or, alternatively, 
to identify specific examples of such correlators whose weak and strong coupling values differ.

In our paper we have not addressed an interesting phenomenon: under
continuous deformations of conformal field theories it is possible for
short multiplets to combine into long ones and to lift from the BPS
bound. By requiring that the spectrum of operators varies
continuously, one can derive certain ``selection rules'' for the types
of states which can combine. These rules can be derived by studying
how the characters of long representations of the superconformal
algebra split into sums of characters of various other
representations, when the former hit the unitarity bound. More
formally these rules can be encoded in the statement that the
"superconformal index" \cite{Kinney:2005ej,Romelsberger:2005eg} of the
theory is invariant under continuous deformations. However, we have
some additional information: the deformation of the theory is
generated by a marginal operator, which is itself a descendant of a
chiral primary. It would be interesting to explore whether this
imposes any additional constraints on the possible combinations of
short multiplets into long ones, besides those imposed by the
superconformal index. We hope to revisit this question in future work.

\section*{Acknowledgments} We would like to thank B. Chowdhury, J. Drummond, M. Guica, N. Lambert, S. Minwalla, S. Raju, S. Ramgoolam for useful discussions. We would like to thank K. Intriligator for useful comments on the draft and especially A. Basu for correspondence.

\appendix

\section{Roots and weights}
\label{sec:groupt}
In this appendix we review some basic facts about Lie algebras in
order to set notation. In every finite dimensional Lie algebra
$\mathfrak{g}$, characterized by a set of hermitian generators
$T_{a}$, there is a maximal subset of commuting generators called
\emph{Cartan subalgebra}, spanned by $H_{i}$, $i = 1,\ldots, m$, where
$m$ is called the \emph{rank} of the algebra.

In a finite-dimensional representation $D$ of the Lie algebra, the
generators are represented by matrices; the Cartan generators can be
simultaneously diagonalized, i.e. we can find a basis of vectors
$\ket{\mu}$ such that \be H_{i} \ket{\mu} = \mu_{i} \ket{\mu} \ee
where the \emph{weight vectors} $\mu$'s are $m$-component vectors with
components $\mu_{i}$. A weight is \emph{positive} if its last non-zero
component is positive and \emph{negative} if its last non-zero
component is negative.\footnote{It is customary to define positive
  weights as having the first non-zero component
  positive. Nevertheless, our definition is more convenient for
  $SU(N)$ groups.} In particular, a weight $\mu^{h}$ such that
$\mu^{h} - \mu$ is positive for every weight $\mu$ is called
\emph{highest weight}. If the representation is irreducible, the
highest weight is unique.

The Lie algebra is a vector space spanned by its generators
$\ket{T_{a}}$, so we can consider the \emph{adjoint representation},
defined by the action of the algebra on itself: \be T_{a} \ket{T_{b}}
= \ket{[T_{a},T_{b}]}  \ee The basis in which the Cartan subalgebra
is diagonal is spanned by $\left\{H_{i},E_{\alpha}\right\}$, and we
have \be [H_{i},H_{j}] = 0, \qquad [H_{i}, E_{\alpha}] = \alpha_{i}
E_{\alpha}, \qquad [E_{\alpha}, E_{-\alpha}] = \alpha \cdot H  \ee
The weights $\alpha$ of the adjoint representation are called
\emph{roots}. A root is called \emph{simple} if it is positive and
cannot be written as a sum of other positive roots. It is possible to
prove that the simple roots are linearly independent and complete, so
the number of simple roots is equal to the rank of the algebra $m$. We
will label the simple roots by $\alpha^{j}$, $j=1,\ldots,m$.

Given an irreducible representation $D$ and a weight $\mu$, the state
$E_{\alpha} \ket{\mu}$ has weight $\mu' = \mu + \alpha$ if $E_{\alpha}
\ket{\mu} \neq 0$. We will refer to the $E_{\alpha^{j}}$ as raising
operators and $E_{-\alpha^{j}} = E_{\alpha^{j}}^{\dagger}$ as lowering
operators.  In particular, the highest weight is annihilated by the
raising operators: \be E_{\alpha^{j}} \ket{\mu^{h}} = 0 \ee since
$\mu^{h} + \alpha$ is not a weight if $\alpha$ is positive. It is
possible to show that \be \frac{2 \alpha^{j} \cdot
  \mu^{h}}{\alpha^{j}\cdot \alpha^{j}} = \ell^{j} \ee where the
$\ell^{j}$ are non-negative integers called \emph{Dynkin
  coefficients}.

It is convenient to introduce a basis of weight vectors $\mu^{j}$ such
that \be \frac{2 \alpha^{j} \cdot \mu^{k}}{\alpha^{j}\cdot \alpha^{j}}
= \delta^{jk} \ee so that the highest weight can be written as
$\mu^{h} = \sum_{j} \ell^{j} \mu^{j}$. The $\mu^{j}$'s are called
\emph{fundamental weights}. Given the highest weight state, all the
states in its irreducible representation can be obtained by acting
with lowering operators: \be E_{-\alpha^{j_{1}}} E_{-\alpha^{j_{2}}}
\cdots E_{-\alpha^{j_{n}}} \ket{\mu^{h}} \ee where $\alpha^{j_{k}}$
are simple roots. The procedure stops when a state of zero norm is
reached. Therefore an irreducible representation is completely
characterized by its highest weight state and can be reconstructed by
acting on this state with lowering operators associated to simple
roots.

As a simple application, notice that if a state has weight $\mu =
\sum_{i} k^{i} \mu^{i}$ with $k^{j} = 0$ for a given $j$, it is
annihilated by the lowering operator $E_{-\alpha^{j}}$. In fact, we
have \be
\label{eqn:zeronorm}
\left<\mu \right| E_{\alpha^{j}} E_{-\alpha^{j}} \ket{\mu} = \alpha^{j} \cdot \mu \left<\mu \right. \ket{\mu} = 0
\ee
so that $E_{-\alpha^{1}} \ket{\mu}$ is a zero-norm state.

\section{1/2 BPS multiplets in ${\cal N}=4$}
\label{sec:su4}

A detailed analysis of the short multiplets in ${\cal N}=4$ can be found in 
\cite{Dolan:2002zh}. Let us start with some group-theoretic elements. The R-symmetry group of the ${\cal N}=4$ algebra in 4 dimensions is $SU(4)$. Its Lie algebra has rank 3, and the Cartan generators are given by:
\begin{equation}
H_{1} = \frac{1}{2}\left(\begin{array}{cccc} 1 & 0 & 0 & 0 \\ 0 & -1 & 0 & 0 \\ 0 & 0 & 0 & 0 \\ 0 & 0 & 0 & 0\end{array}\right), \;
H_{2} = \frac{1}{\sqrt{12}}\left(\begin{array}{cccc} 1 & 0 & 0 & 0 \\ 0 & 1 & 0 & 0 \\ 0 & 0 & -2 & 0 \\ 0 & 0 & 0 & 0\end{array}\right), \;
H_{3} = \frac{1}{\sqrt{24}}\left(\begin{array}{cccc} 1 & 0 & 0 & 0 \\ 0 & 1 & 0 & 0 \\ 0 & 0 & 1 & 0 \\ 0 & 0 & 0 & -3\end{array}\right)
\end{equation}
The weights of the fundamental representation are given by
\begin{equation}
\label{eqn:definingweights}
v^{1} = \begin{pmatrix} \frac{1}{2} \\[5pt] \frac{1}{\sqrt{12}} \\[5pt] \frac{1}{\sqrt{24}} \end{pmatrix}, \quad
v^{2} = \begin{pmatrix} -\frac{1}{2} \\[5pt] \frac{1}{\sqrt{12}} \\[5pt] \frac{1}{\sqrt{24}} \end{pmatrix}, \quad
v^{3} = \begin{pmatrix} 0 \\[5pt] -\frac{2}{\sqrt{12}} \\[5pt] \frac{1}{\sqrt{24}} \end{pmatrix}, \quad
v^{4} = \begin{pmatrix} 0 \\[5pt] 0 \\[5pt] -\frac{3}{\sqrt{24}} \end{pmatrix}
\end{equation}
the roots by
\begin{equation}
\alpha^{1} = v^{1} - v^{2} = \begin{pmatrix} 1 \\[5pt] 0 \\[5pt] 0 \end{pmatrix}, \quad
\alpha^{2} = v^{2} - v^{3} = \begin{pmatrix} -\frac{1}{2} \\[5pt] \frac{\sqrt{3}}{2} \\[5pt] 0 \end{pmatrix}, \quad 
\alpha^{3} = v^{3} - v^{4} = \begin{pmatrix} 0 \\[5pt] -\frac{1}{\sqrt{3}} \\[5pt] \frac{2}{\sqrt{6}} \end{pmatrix}
\end{equation}
and the fundamental weights by
\begin{equation}
\mu^{1} = v^{1} = \begin{pmatrix} \frac{1}{2} \\[5pt] \frac{1}{\sqrt{12}} \\[5pt] \frac{1}{\sqrt{24}} \end{pmatrix}, \quad
\mu^{2} = v^{1} + v^{2} = \begin{pmatrix} 0 \\[5pt] \frac{1}{\sqrt{3}} \\[5pt] \frac{1}{\sqrt{6}} \end{pmatrix}, \quad
\mu^{3} = v^{1} + v^{2} + v^{3} = \begin{pmatrix} 0 \\[5pt] 0 \\[5pt] \frac{\sqrt{3}}{2\sqrt{2}} \end{pmatrix} \quad
\end{equation}
so that $\frac{2 \alpha^{j} \cdot \mu^{k}}{\alpha^{j} \cdot \alpha^{j}} = \delta^{jk}$. Every irreducible representation is uniquely characterized by the Dynkin label $[k_{1},k_{2},k_{3}]$, meaning that the highest weight is $\mu^{h} = k_{1} \mu^{1} + k_{2} \mu^{2} + k_{3} \mu^{3}$. The complex conjugate of the representation $[k_{1},k_{2},k_{3}]$ is $[k_{3},k_{2},k_{1}]$.

We will denote the raising operators $E_{\alpha^{1}}$, $E_{\alpha^{2}}$ and $E_{\alpha^{3}}$ by $E_{1}$, $E_{2}$ and $E_{3}$ respectively, and the corresponding lowering operators by $E_1^\dagger$, $E_2^\dagger$ and $E_3^\dagger$.

The highest weight for the fundamental representation is $v^{1} = \mu^{1}$, therefore the Dynkin label is simply $[1,0,0]$. Sometimes it is convenient to denote representations by their dimension $\mathbf{d}$, so that the fundamental representation $[1,0,0]$ is denoted by $\mathbf{4}$ and its complex conjugate $[0,0,1]$ by $\bar{\mathbf{4}}$. Finally, the six-dimensional representation $[0,1,0]$, or $\mathbf{6}$, corresponds to the fundamental representation of $SO(6)$ through the local isomorphism $SO(6)$ $\approx$ $SU(4)$.

Representations $[k_{1},k_{2},k_{3}]$ are conveniently represented in terms of Young tableaux with $k_{3}$ columns with 3 boxes, $k_{2}$ columns with 2 boxes and $k_{1}$ columns with 1 box:

\begin{equation}
\begin{ytableau}
\, & \none[\dots] &  &  & \none[\dots] & & & \none[\dots] & \\
\, & \none[\dots] &  &  & \none[\dots] &  \\
\, & \none[\dots] &   \\
\end{ytableau}
\end{equation}
In particular, the fundamental representation $[1,0,0]$ is denoted by
\begin{equation}
\ydiagram{1}
\end{equation}
and the representations $[0,k,0]$ by a Young tableau with $2k$ boxes:

\begin{equation}
\overbrace{
\begin{ytableau}
\, & & \none[\dots] &  \\
\, & & \none[\dots] &   \\
\end{ytableau}
}^{k \text{ times}}
\end{equation}
We refer to \cite{Georgi:1982jb} for more details.
\subsection{The $[0,k,0]$ multiplet}
\label{sec:oko}
The representations of the form $[0,k,0]$ are of particular importance, since the 1/2 BPS multiplets $\phi^{(k,\vec{m})}$ in the $\mathcal{N} = 4$ theory sit in such representations. The vector $\vec{m}$ denotes the weight associated to a particular state in the representation. The highest and lowest weight states are denoted respectively by $\phi^{(k,+)}$ and $\phi^{(k,-)}$.

These representations can be constructed by taking tensor products of $k$ $[0,1,0]$ representations. The $[0,1,0]$, or $\mathbf{6}$, representation can be obtained as the antisymmetric product of two $\mathbf{4}$ representations. It is usually more convenient to work with a $SO(6)$ notation $\phi^{i}$, $i=1,\ldots,6$. The six scalar fields of $\mathcal{N}=4$ super Yang--Mills sit in this representation. The irreducible representations $[0,k,0]$ for the chiral primaries correspond to traceless symmetric tensors $C_{i_{1}\ldots i_{k}}$:
\be
C_{i_{1}\ldots i_{n}} \mathrm{Tr}(\phi^{i_{1}} \cdots \phi^{i_{k}})
\ee
where the trace is over the $SU(N)$ gauge group. The highest weight state in this notation is
\be
\mathrm{Tr}\left(Z^{k}\right) = \mathrm{Tr}\left((\phi^{1} + i \phi^{2})^{k}\right)
\ee
while the lowest weight is
\be
\mathrm{Tr}\left(\bar{Z}^{k}\right) = \mathrm{Tr}\left((\phi^{1} - i \phi^{2})^{k}\right)
\ee

The left-chiral supercharges ${\bf Q}$ sit in the fundamental representation of $SU(4)$, and we will use a basis ${\bf Q}^{i}$, $i=1,\ldots,4$ corresponding to the weights $v^{i}$, $i=1,\ldots,4$ defined in equation \eqref{eqn:definingweights} (in this section we ignore the Lorentz indices).

When we act with ${\bf Q}$ on $\phi$ we obtain a tensor product representation that can be decomposed as the sum of two irreducible representations as follows
\begin{equation}
\begin{ytableau}
\, & & \none[\dots] &  \\
\, & & \none[\dots] &   \\
\end{ytableau}\; \bigotimes \;
\ydiagram{1}\; = \; \begin{ytableau}
\, & & \none[\dots] & & \\
\, & & \none[\dots] &   \\
\end{ytableau}\; \bigoplus \;
\begin{ytableau}
\, & & \none[\dots] &  \\
\, & & \none[\dots] &   \\
\,
\end{ytableau}
\end{equation}
or
\begin{equation}
[1,0,0] \otimes [0,k,0] = [1,k,0] \oplus [0,k-1,1]
\end{equation}

Using the $\mathcal{N} = 4$ algebra and the condition $\Delta = k$, it is easy to see that the highest weight in $[1,k,0]$, namely $[{\bf Q}^1 ,\phi^{(k,+)}]$, has zero norm. Furthermore, from equation \eqref{eqn:zeronorm} we have $[E_{1}^{\dagger}, \phi^{(k,+)}] = 0$, which means that:
\begin{equation}
[E_{1}^{\dagger}, [{\bf Q}^1, \phi^{(k,+)}]] = [[E_{1}^{\dagger}, {\bf Q}^1], \phi^{(k,+)}] = [{\bf Q}^2 ,\phi^{(k,+)}]
\end{equation}
Therefore $[{\bf Q}^2, \phi^{(k,+)}]$ belongs to the null representation as well, being a descendant of the highest weight $[{\bf Q}^1 ,\phi^{(k,+)}]$. Therefore we will write
\be
[{\bf Q}^1, \phi^{(k,+)} ]= 0, \qquad [{\bf Q}^2, \phi^{(k,+)}] = 0
\ee
Analogously, we have
\be
[{\bf Q}^3, \phi^{(k,-)}] = 0, \qquad [{\bf Q}^4, \phi^{(k,-)}] = 0
\ee

Finally, notice that the decomposition of $[k_{1},k_{2},k_{3}] \otimes [k'_{1},k'_{2},k'_{3}]$ into a sum of irreducible representations contains the trivial representation if and only if $[k'_{1},k'_{2},k'_{3}]$ is the complex conjugate representation of $[k_{1},k_{2},k_{3}]$, that is $[k_{3},k_{2},k_{1}]$. In particular, the tensor product $[0,k,0] \otimes R$, where $R$ is an arbitrary (not necessarily irreducible) representation, contains the trivial representation if and only if $R$ contains the representation $[0,k,0]$.
\subsection{The $[0,2,0]$ multiplet}
\label{sec:oto}
We summarize some (well known) facts about the $[0,2,0]$ 1/2 BPS multiplet of ${\cal N}=4$ SYM. This multiplet is special because it contains the conserved currents and also the marginal operators. 

The highest weight of the multiplet is the operator ${\rm Tr}(Z^2)$, where $Z=\phi^1+i \phi^2$. This operator is annihilated by 1/2 of the left chiral and 1/2 of the right chiral supercharges. Here we use the notation ${\bf Q}^i_a$, $\overline{\bf Q}_{j,\dot{a}}$ where $i,j$ are $SU(4)$ indices and $a,\dot{a}$ are $(1/2,0)$ and $(0,1/2)$ Lorentz spinor indices. The operator ${\rm Tr}(Z^2)$ is annihilated by the left chiral ${\bf Q}^1_{a},{\bf Q}^2_{a}$ and the right chiral $\overline{\bf Q}_{3,\dot{a}}, \overline{\bf Q}_{4,\dot{a}}$, and is not annihilated by the rest of the supercharges.

Let us consider the four left chiral supercharges which do not annihilate the operator ${\rm Tr}(Z^2)$, namely ${\bf Q}^3_{a},{\bf Q}^4_{a}$ where the spinor indices can be $a=1,2$. We notice that according to the ${\cal N}=4$ superconformal algebra, these operators anticommute among themselves. Hence if we consider a nested (anti)-commutator of these supercharges, then the order in which the supercharges appear is not important and we can bring them to any desired order. The marginal operator ${\cal O}_\tau$ can then be written as
\be
\label{hola}
{\cal O}_\tau = \{ {\bf Q}^4_{1},[{\bf Q}^4_{2},\{{\bf Q}^3_{1},[{\bf Q}^3_{2},{\rm Tr}(Z^2)]\}]\}
\ee
It is straightforward to check using the superconformal algebra that this operator is Lorentz scalar, conformal primary and has $\Delta =4$. Similarly, if we act on it with the four right chiral supercharges which do not annihilate it we get the conjugate marginal opeator
\be
\label{antihola}
\overline{\cal O}_\tau = \{\overline{\bf Q}_{2,\dot{1}},[\overline{\bf Q}_{2,\dot{2}},\{\overline{\bf Q}_{1,\dot{1}},[\overline{\bf Q}_{1,\dot{2}},{\rm Tr}(Z^2)]\}]\}
\ee

Similar statements hold for the conjugate operator ${\rm Tr}(\overline{Z}^2)$, which is the $SU(4)$ lowest weight state of the $[0,2,0]$ multiplet. This operator is also annihilated by 1/2 of the left chiral and 1/2 of the right chiral supercharges --- more specifically it is annihilated by ${\bf Q}^3_a, {\bf Q}^4_a$ and $\overline{\bf Q}_{1,\dot{a}},\overline{\bf Q}_{2,\dot{a}}$. If we act on it with the four left chiral supercharges which do not annihilate it we have
\be
\label{holb}
{\cal O}_\tau = \{ {\bf Q}^2_1,[ {\bf Q}^2_2,\{{\bf Q}^1_1,[{\bf Q}^1_2,{\rm Tr}(\overline{Z}^2)]\}]\}
\ee
while acting with the right chiral supercharges
\be
\label{antiholb}
\overline{\cal O}_\tau = \{\overline{\bf Q}_{4,\dot{1}},[\overline{\bf Q}_{4,\dot{2}},\{\overline{\bf Q}_{3,\dot{1}},[\overline{\bf Q}_{3,\dot{2}},{\rm Tr}(\overline{Z}^2)]\}]\}
\ee

The expressions \eqref{hola} and \eqref{antiholb} are manifestly related by complex conjugation. On the other hand, the fact that ${\cal O}_\tau$ (and similarly $\overline{\cal O}_\tau$) can be written either as \eqref{hola} or \eqref{holb} is less obvious and special to ${\cal N}=4$ theories.

The reason that we went into such a detailed presentation here is because the marginal operators in the ${\cal N}=4$ have some special properties, which differ from those encountered in theories with less supersymmetry. If we think of the operator ${\rm Tr}(Z^2)$ as a ``chiral primary'' and that of ${\rm Tr}(\overline{Z}^2)$ as an ``anti-chiral'', we notice that both the holomorphic ${\cal O}_\tau$ and antiholomorphic $\overline{\cal O}_\tau$ marginal operators can be written as descendant of either the chiral or the anti-chiral primary. This is in contrast to what happens in less supersymmetric theories, where the holomorphic deformations are paired with descendants of chiral primaries and anti-holomorphic with descendants of anti-chiral. 

Similar special properties of marginal operators are encountered in 2d ${\cal N}=(4,4)$ theories, as explained in detail in \cite{arXiv:0809.0507}.

\section{Null states and short multiplets}
\label{sec:nullconditions}
In this appendix we prove the null conditions \eqref{descen} and \eqref{nullstate}. The proof is very similar in both cases, and we begin with the two-dimensional case which is technically simpler.
\subsection{Structure of null conditions in ${\cal N}=(4,4)$}
For simplicity we drop all extra indices/boldface notation and denote the supercharges by $G^\pm \equiv {\bf G}_{-\frac12}^{\pm r}$, $J\equiv J^-$ and $\phi=\phi^{(j,j)}$,
i.e. the highest weight state in the (short) representation. Also for simplicity we assume that the highest weight state is bosonic (if fermionic some commutators have to be replaced by anticommutators). By definition we have 
$[G^+,\phi]=0$. What we want to prove is that
$$
[G^+, \overbrace{[J,\ldots[J,}^n \phi]\ldots] \sim [G^-, \overbrace{[J,\ldots[J,}^{n-1} \phi]\ldots]
$$
We will prove it recursively. For $n=1$ we have
$$
[G^+, [J, \phi]] = [[G^+,J],\phi] + [J ,[G^+,\phi] ]= [G^-, \phi]
$$
where we used that the second term is zero and the algebra relation $[G^+,J]=G^-$.

Next, let us assume that the condition is true for $n$ and show that it also true for $n+1$. We have
$$
[G^+, \overbrace{[J,\ldots[J,}^{n+1} \phi]\ldots]   =
 [[G^+,J], \overbrace{[J,\ldots[J,}^n \phi]\ldots]  + [J,[ G^+, \overbrace{[J,\ldots[J,}^n \phi]\ldots]
$$
$$ 
 = [G^-,\overbrace{[J,\ldots[J,}^n \phi]\ldots]   + [J,[ G^-,\overbrace{[J,\ldots[J,}^{n-1} \phi]\ldots] 
$$
To get this we used the algebra $[G^+,J]=G^-$ and the inductive hypothesis. Now we commute $G^{-}$ to the left and we have
$$
[G^+, \overbrace{[J,\ldots[J,}^{n+1} \phi]\ldots] = 
[G^-, \overbrace{[J,\ldots[J,}^n \phi]\ldots]+[[J,G^{-}], \overbrace{[J,\ldots[J,}^{n-1} \phi]\ldots]
$$
Now from the algebra we have $[J,G^-]=0$, so we have proved the desired relation.

\subsection{Structure of null conditions for ${\cal N}=4$}

We now move to the four dimensional case where we want to prove \eqref{nullstate}, which reads
\begin{equation}
\label{nullstateb}
[{\bf Q}^\star,\phi_I^{(k_1,\vec{m}_1)}] = \sum_{j\neq\star}[ {\bf Q}^j ,{\cal X}_j]
\end{equation}
here all supercharges are left chiral. We have chosen a basis of
supercharges that have definite weight under the Cartan
subalgebra. Let us consider one of the supercharges that annihilate a
highest weight state $\phi^{(k,+)}$ (namely either ${\bf Q}^{1}$ or ${\bf Q}^{2}$)
and call it ${\bf Q}^\star$. Hence we have
$$
[{\bf Q}^\star,\phi^{(k,+)}]=0
$$
In this case equation \eqref{nullstateb} is trivially satisfied.

Let us prove equation \eqref{nullstateb} in the case where the operator is the first $SU(4)$ ``descendant" i.e. $[E_i^\dagger, \phi^{(k,+)}]$. We
have
$$
[{\bf Q}^\star, [E_i^\dagger, \phi^{(k,+)}]] = [E_i^\dagger ,[{\bf Q}^\star ,\phi^{(k,+)}]] + [[{\bf Q}^\star,E_i^\dagger],\phi^{(k,+)}]= [{\bf Q}' , \phi^{(k,+)}]
$$
The first term is zero while the term $[{\bf Q}^\star,E_i^\dagger]= {\bf Q}'$
is another supercharge. However the important point is that the SU(4) weight of
the supercharge ${\bf Q}'$ is equal to the weight
of ${\bf Q}^*$ minus the root $\alpha_{i}$, so definitely ${\bf Q}'\neq {\bf Q}^\star$. Hence \eqref{nullstateb} is proven in this case.

In general, let us assume that the relation is true for an $n$ descendant, that is
\be
\label{inddd}
[{\bf Q}^{\star },[E_{i_{1}}^{\dagger},[ \ldots ,[E_{i_{n}}^{\dagger} , \phi^{(k,+)}]\ldots] = \sum_{i\neq \star} [{\bf Q}^{i} ,\phi^{(k,\vec{m}_{i})}]
\ee
where the weight of each ${\bf Q}^{i}$ is strictly smaller than that of ${\bf Q}^{\star}$. We now show that the relation holds for an $n+1$ descendant as well. We have
$$
[{\bf Q}^\star ,[E_{i}^{\dagger},[ E_{i_{1}}^{\dagger}, \ldots [E_{i_{n}}^{\dagger},  \phi^{(k,+)}]\ldots] = [E_{i}^{\dagger},[ {\bf Q}^{\star},[ E_{i_{1}}^{\dagger}, \ldots [E_{i_{n}}^{\dagger}, \phi^{(k,+)}]
\ldots]
$$
$$
 + [[{\bf Q}^{\star}, E_{i}^{\dagger}] , [E_{i_{1}}^{\dagger} ,\ldots[ E_{i_{n}}^{\dagger} , \phi^{(k,+)}]\ldots]
$$
By using the inductive hypothesis \eqref{inddd} on the right hand side, we have
\be
[{\bf Q}^\star ,[E_{i}^{\dagger} ,[E_{i_{1}}^{\dagger} \ldots ,[E_{i_{n}}^{\dagger}  ,\phi^{(k,+)}]\ldots] = [E_{i}^{\dagger} ,\sum_{j\neq \star} [{\bf Q}^{j}, \phi^{(k,\vec{m}_{j})}]] + [{\bf Q}',[ E_{i_{1}}^{\dagger} \ldots [E_{i_{n}}^{\dagger}  ,\phi^{(k,+)}]\ldots]
\ee
where the weight of ${\bf Q}' \equiv  [{\bf Q}^{\star}, E_{i}^{\dagger}]$ is strictly smaller than the weight of ${\bf Q}^{\star}$. A further manipulation gives
$$
[{\bf Q}^\star,[ E_{i}^{\dagger},[ E_{i_{1}}^{\dagger} \ldots [E_{i_{n}}^{\dagger}  ,\phi^{(k,+)}]\ldots] = \sum_{j\neq \star} [{\bf Q}^{j},[ E_{i}^{\dagger}, \phi^{(k,\vec{m}_{j})}]] 
$$
$$+  \sum_{j\neq \star} [[E_{i}^{\dagger}, {\bf Q}^{j}] ,\phi^{(k,\vec{m}_{j})}] + [{\bf Q}',[ E_{i_{1}}^{\dagger} \ldots [E_{i_{n}}^{\dagger} , \phi^{(k,+)}]\ldots]
$$
and since ${\bf Q}'' \equiv [E_{i}^{\dagger}, {\bf Q}^j]$ has a smaller weight than ${\bf Q}^j$, we have proved the desired relation. It is trivial to repeat the above steps for ${\bf Q}^3$ and ${\bf Q}^4$ by starting with the lowest weight $\phi^{(k,-)}$ and working ``upwards''.

\bibliography{bibliography}

\begin{thebibliography}{10}

\bibitem{Maldacena:1997re}
J.~M. Maldacena, ``{The Large N limit of superconformal field theories and
  supergravity},'' {\em Adv.Theor.Math.Phys.}, vol.~2, pp.~231--252, 1998,
  hep-th/9711200.

\bibitem{hep-th/9802109}
S.~Gubser, I.~R. Klebanov, and A.~M. Polyakov, ``{Gauge theory correlators from
  noncritical string theory},'' {\em Phys.Lett.}, vol.~B428, pp.~105--114,
  1998, hep-th/9802109.

\bibitem{hep-th/9802150}
E.~Witten, ``{Anti-de Sitter space and holography},'' {\em
  Adv.Theor.Math.Phys.}, vol.~2, pp.~253--291, 1998, hep-th/9802150.

\bibitem{hep-th/9806074}
S.~Lee, S.~Minwalla, M.~Rangamani, and N.~Seiberg, ``{Three point functions of
  chiral operators in D = 4, N=4 SYM at large N},'' {\em Adv.Theor.Math.Phys.},
  vol.~2, pp.~697--718, 1998, hep-th/9806074.

\bibitem{D'Hoker:1998tz}
E.~D'Hoker, D.~Z. Freedman, and W.~Skiba, ``{Field theory tests for correlators
  in the AdS / CFT correspondence},'' {\em Phys.Rev.}, vol.~D59, p.~045008,
  1999, hep-th/9807098.

\bibitem{D'Hoker:1999ea}
E.~D'Hoker, D.~Z. Freedman, S.~D. Mathur, A.~Matusis, and L.~Rastelli,
  ``{Extremal correlators in the AdS / CFT correspondence},'' 1999,
  hep-th/9908160.

\bibitem{hep-th/0703001}
M.~R. Gaberdiel and I.~Kirsch, ``{Worldsheet correlators in AdS(3)/CFT(2)},''
  {\em JHEP}, vol.~0704, p.~050, 2007, hep-th/0703001.

\bibitem{hep-th/0703022}
A.~Dabholkar and A.~Pakman, ``{Exact chiral ring of AdS(3) / CFT(2)},'' {\em
  Adv.Theor.Math.Phys.}, vol.~13, pp.~409--462, 2009, hep-th/0703022.

\bibitem{Pakman:2007hn}
A.~Pakman and A.~Sever, ``{Exact N=4 correlators of AdS(3)/CFT(2)},'' {\em
  Phys.Lett.}, vol.~B652, pp.~60--62, 2007, 0704.3040.

\bibitem{Taylor:2007hs}
M.~Taylor, ``{Matching of correlators in AdS(3) / CFT(2)},'' {\em JHEP},
  vol.~0806, p.~010, 2008, 0709.1838.

\bibitem{Intriligator:1998ig}
K.~A. Intriligator, ``{Bonus symmetries of N=4 superYang-Mills correlation
  functions via AdS duality},'' {\em Nucl.Phys.}, vol.~B551, pp.~575--600,
  1999, hep-th/9811047.

\bibitem{Intriligator:1999ff}
K.~A. Intriligator and W.~Skiba, ``{Bonus symmetry and the operator product
  expansion of N=4 SuperYang-Mills},'' {\em Nucl.Phys.}, vol.~B559,
  pp.~165--183, 1999, hep-th/9905020.

\bibitem{Eden:1999gh}
B.~Eden, P.~S. Howe, and P.~C. West, ``{Nilpotent invariants in N=4 SYM},''
  {\em Phys.Lett.}, vol.~B463, pp.~19--26, 1999, hep-th/9905085.

\bibitem{Petkou:1999fv}
A.~Petkou and K.~Skenderis, ``{A Nonrenormalization theorem for conformal
  anomalies},'' {\em Nucl.Phys.}, vol.~B561, pp.~100--116, 1999,
  hep-th/9906030.

\bibitem{Howe:1999hz}
P.~S. Howe, C.~Schubert, E.~Sokatchev, and P.~C. West, ``{Explicit construction
  of nilpotent covariants in N=4 SYM},'' {\em Nucl.Phys.}, vol.~B571,
  pp.~71--90, 2000, hep-th/9910011.

\bibitem{Heslop:2001gp}
P.~Heslop and P.~S. Howe, ``{OPEs and three-point correlators of protected
  operators in N=4 SYM},'' {\em Nucl.Phys.}, vol.~B626, pp.~265--286, 2002,
  hep-th/0107212.

\bibitem{Basu:2004nt}
A.~Basu, M.~B. Green, and S.~Sethi, ``{Some systematics of the coupling
  constant dependence of N=4 Yang-Mills},'' {\em JHEP}, vol.~0409, p.~045,
  2004, hep-th/0406231.

\bibitem{arXiv:0809.0507}
J.~de~Boer, J.~Manschot, K.~Papadodimas, and E.~Verlinde, ``{The Chiral ring of
  AdS(3)/CFT(2) and the attractor mechanism},'' {\em JHEP}, vol.~0903, p.~030,
  2009, 0809.0507.

\bibitem{hep-th/9304053}
K.~Ranganathan, H.~Sonoda, and B.~Zwiebach, ``{Connections on the state space
  over conformal field theories},'' {\em Nucl.Phys.}, vol.~B414, pp.~405--460,
  1994, hep-th/9304053.

\bibitem{arXiv:0910.4963}
K.~Papadodimas, ``{Topological Anti-Topological Fusion in Four-Dimensional
  Superconformal Field Theories},'' {\em JHEP}, vol.~1008, p.~118, 2010,
  0910.4963.

\bibitem{Seiberg:1988pf}
N.~Seiberg, ``{Observations on the Moduli Space of Superconformal Field
  Theories},'' {\em Nucl.Phys.}, vol.~B303, p.~286, 1988.

\bibitem{Kutasov:1988xb}
D.~Kutasov, ``{Geometry On The Space Of Conformal Field Theories And Contact
  Terms},'' {\em Phys.Lett.}, vol.~B220, p.~153, 1989.

\bibitem{Cardona:2010qf}
C.~A. Cardona and I.~Kirsch, ``{Worldsheet four-point functions in
  $AdS_3/CFT_2$},'' {\em JHEP}, vol.~1101, p.~015, 2011, 1007.2720.

\bibitem{Kirsch:2011na}
I.~Kirsch and T.~Wirtz, ``{Worldsheet operator product expansions and p-point
  functions in $AdS_3/CFT_2$},'' {\em JHEP}, vol.~1110, p.~049, 2011,
  1106.5876.

\bibitem{Strominger:1996sh}
A.~Strominger and C.~Vafa, ``{Microscopic origin of the Bekenstein-Hawking
  entropy},'' {\em Phys.Lett.}, vol.~B379, pp.~99--104, 1996, hep-th/9601029.

\bibitem{Maldacena:1997de}
J.~M. Maldacena, A.~Strominger, and E.~Witten, ``{Black hole entropy in M
  theory},'' {\em JHEP}, vol.~9712, p.~002, 1997, hep-th/9711053.

\bibitem{Minasian:1999qn}
R.~Minasian, G.~W. Moore, and D.~Tsimpis, ``{Calabi-Yau black holes and (0,4)
  sigma models},'' {\em Commun.Math.Phys.}, vol.~209, pp.~325--352, 2000,
  hep-th/9904217.

\bibitem{Dolan:2002zh}
F.~Dolan and H.~Osborn, ``{On short and semi-short representations for
  four-dimensional superconformal symmetry},'' {\em Annals Phys.}, vol.~307,
  pp.~41--89, 2003, hep-th/0209056.

\bibitem{Kinney:2005ej}
J.~Kinney, J.~M. Maldacena, S.~Minwalla, and S.~Raju, ``{An Index for 4
  dimensional super conformal theories},'' {\em Commun.Math.Phys.}, vol.~275,
  pp.~209--254, 2007, hep-th/0510251.

\bibitem{Romelsberger:2005eg}
C.~Romelsberger, ``{Counting chiral primaries in N = 1, d=4 superconformal
  field theories},'' {\em Nucl.Phys.}, vol.~B747, pp.~329--353, 2006,
  hep-th/0510060.

\bibitem{Georgi:1982jb}
H.~Georgi, ``{Lie Algebras In Particle Physics. From Isospin To Unified
  Theories},'' {\em Front.Phys.}, vol.~54, pp.~1--255, 1982.

\end{thebibliography}
\bibliographystyle{hieeetr}
\end{document}